\documentclass[12pt,preprint2]{aastex}

\shorttitle{Kinematics of Sa2-237}
\shortauthors{Schwarz et al.}

\begin{document}

\title{Kinematic and morphological modeling of the bipolar nebula 
Sa2-237\altaffilmark{1}}

\author{Hugo E. Schwarz}
\affil{Cerro Tololo Inter-American Observatory, NOAO\altaffilmark{2}, 
Casilla 603, La Serena, Chile}
\email{hschwarz@ctio.noao.edu}

\author{Romano R. L. Corradi}
\affil{ING, Apartado 321, E-38700 Sta. Cruz de La Palma, Espa\~{n}a}
\email{rcorradi@ing.iac.es}

\author{Rodolfo Montez Jr.\altaffilmark{3}}
\affil{Dept. of Astronomy, University of Texas at Austin, USA}
\email{rudy@astro.as.utexas.edu}

\altaffiltext{1}{Based
on observations collected at the European Southern Observatory,
Chile}

\altaffiltext{2}{Cerro Tololo Inter-American Observatory, National Optical 
Astronomy Observatories, operated by the Association of Universities
for Research in Astronomy, Inc., under a cooperative agreement with
the National Science Foundation.}

\altaffiltext{3}{Cerro Tololo Inter-American Observatory Research 
Experiences for Undergraduates (REU) Program student.}

\begin{abstract}

We present [OIII]500.7nm and H$\alpha$+[NII] images and long-slit,
high resolution echelle spectra in the same spectral regions of
Sa2--237, a possible bipolar planetary nebula. The image shows a
bipolar nebula of about 34\arcsec extent, with a narrow waist, and
showing strong point symmetry about the central object, indicating
it's likely binary nature. The long slit spectra were taken over the
long axis of the nebula, and show a distinct ``eight'' shaped pattern
in the velocity--space plot, and a maximum projected outflow velocity
of V$_{exp}=$106 km$\cdot$s$^{-1}$, both typical of expanding bipolar
planetary nebulae. By model fitting the shape and spectrum of the
nebula simultaneously, we derive the inclination of the long axis to
be 70$^o$, and the maximum space velocity of expansion to be $\leq$308
km$\cdot$s$^{-1}$. Due to asymmetries in the velocities we adopt a new
value for the system's heliocentric radial velocity of -30km.s$^{-1}$.
We use the IRAS \& 21cm radio fluxes, the energy distribution, and the
projected size of Sa2-237 to estimate it's distance to be
2.1$\pm$0.37kpc. At this distance Sa2-237 has a luminosity of 340
L$_{\sun}$, a size of 0.37pc, and -- assuming constant expansion
velocity -- a nebular age of 624 years. The above radial velocity
\& distance place Sa2--237 in the disk of the Galaxy at z=255pc, albeit with
somewhat peculiar kinematics.

\end{abstract}


\section{Introduction}

Most low mass stars spend some time at the end of their lives as
Planetary Nebulae (PNe) before they die away as cooling white
dwarfs. The processes by which these spherically symmetrical low mass
stars become --often highly asymmetric-- PNe are not well understood.
Especially strong asymmetries exist in the sub--class of bipolar PNe.
Candidates for the formation of bipolar PNe include binary central
objects, fast rotating single stars with strong magnetic fields, and
interacting wind models.  Although, for the more extremely collimated
nebulae, a binary system is becoming a generally accepted necessity.
A glance at Balick (1987), Schwarz et al. (1992) and G\'{o}rny et
al. (1999) shows that a majority of PNe are significantly
asymmetrical.  A recent overview of asymmetrical PNe is given in
Kastner et al. (2000).

Bipolar PNe have been shown to have properties very different from
those of the general population of PNe (Corradi \& Schwarz, 1995).
They form an interesting group combining shocks and photoionization,
dust and gas, and very high outflow velocities with interesting
kinematics.  They have giant dimensions (several are $>$ 1pc), central
stars with higher mass (as in type I PNe), and surprisingly, evidence
shows they have lower than typical luminosities (Corradi \& Schwarz,
1995).  In some cases, the combination of high temperatures and
excitation with low luminosity and the presence of a variable Near
Infra-Red (NIR) source indicates the presence of a compact, hot binary
component. In the case of M2--9, measurable expansion parallax
(Schwarz et al. 1997) and internal movements (Doyle et al. 2000) have
been observed.  Several bipolar PNe possess point--symmetry and are,
strictly speaking, not plane--symmetric as are the other bipolar
PNe. The point--symmetry can be spectacularly clear (Schwarz 1993,
L\'{o}pez et al. 1993) and has been suggested to be due to precession
in the binary orbit of the central object (Schwarz et al. 1992, Livio
2000).

The bipolar nebula Sa2-237 (PN G011.1+07.0, 17$^h$44$^m$42$^s$
-15\degr45\arcmin13\arcsec J2000.)  which is briefly presented in a
poster paper by Masegosa et al. (1999), is discussed in some detail in
this paper. Information on Sa2-237 is sparse, a SIMBAD search only
resulted in IRAS fluxes, and a heliocentric systemic radial
velocity measurement of -81km$\cdot$s$^{-1}$.  The 18 references since 1983
are all to catalogs or papers discussing samples of PNe; apart from
the Masegosa poster, no papers have been found about the object
itself.  The kinematic and morphological analysis of Sa2-237
presented here will hopefully advance the understanding of the interesting
characteristics of bipolar PNe.

\section{Observations \& Data}

The spectra were secured with EMMI, the ESO Multi Mode Instrument,
mounted at the NTT, during the night of 2000.12.04. Using the long
slit echelle mode with grating 14 results in a resolution of 0.004nm
per pixel on the CCD. A slit width of 0\farcs8 on the sky gives a
spectral resolving power of 70000 spatially projecting onto 2.2
pixels. The CCD was a Tektronix TK2048EB 2k$^2$ chip with 24$\mu$
pixels each covering 0\farcs28 on the sky along the slit of 300\arcsec 
length. The CCD has a RON of 5e, a dark current of 1.7e
pix$^{-1}$$\cdot$hr$^{-1}$, and was used with a gain of
2.17e$\cdot$ADU$^{-1}$ in the unbinned mode.  The order of the echelle
grating was selected by using an H$\alpha$ filter (ESO No.596) passing
also the [NII] lines at 654.8
\& 658.3nm, or an [OIII]500.7nm filter (ESO No.589). The exposure times were 
4800s (in two 2400s exposures) for the H$\alpha$ spectrum, and 2400s for
the [OIII] spectrum. For details on the NTT, EMMI and it's filters and
gratings, see \url{www.ls.eso.org/lasilla/Telescopes/NEWNTT/}

The images were also taken with EMMI, but used in the RILD imaging
mode. The same ESO filters used for the spectra were used to acquire
the images, and the exposure times were 500s for the [OIII] image and
300s for H$\alpha$.

The observations were reduced in the usual way using the IRAF package
(see \url{www.iraf.noao.edu}) and the resulting images and spectra are
shown in Figures~1 \& 2 respectively.

\begin{figure*}
\epsscale{2.0}
\plotone{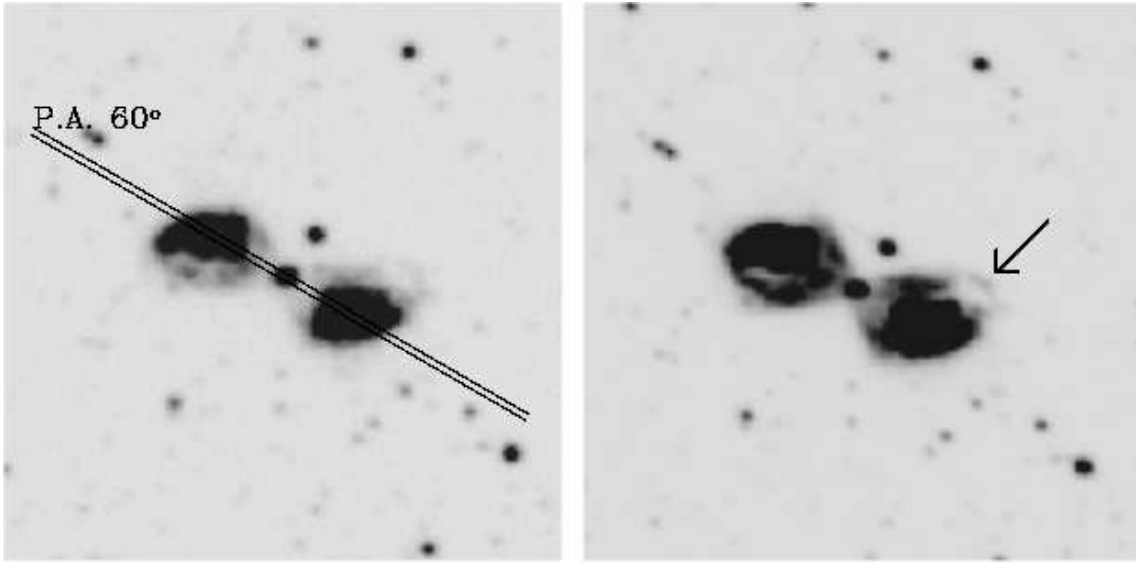}
\caption{The [OIII]500.7 (left) and H$\alpha$+[NII]658.3 (right) images of 
Sa2--237. The position of the slit is indicated in the [OIII] image,
and it's width of 0\farcs8 is shown to scale. The arrow indicates the
faint loop discussed in the text. Each image is 70\arcsec on a side,
with N~$\uparrow$ and E~$\leftarrow$.\label{fig1}}
\end{figure*}

Our images show a bipolar nebula with a narrow waist extending over
about 34\arcsec, and point symmetry about the central object. The
lobes are oriented more or less to the E and W, and are brighter in
the northern part of the E lobe and the southern part of the W
lobe. In the extreme NW of the W lobe there is a faint loop visible,
indicated by an arrow in Figure~1, which may have an even fainter
counterpart in the opposite lobe. Deeper imaging may be able to
resolve this, and also determine if there is fainter emission further
away from the central object. Several other bipolars have had faint
emission detected far from the brighter parts of the nebula: e.g.
MyCn18 (Bryce et al. 1997) \& M2--9 Kohoutek \& Surdej 1980) therefore
it is worth looking for this around Sa2--237. The object is extended
in 2MASS NIR images, with the E lobe being the brighter, and the central
object is much brighter than the nebula.
 
The position of the slit used to acquire the spectra in Figure~2 is
over plotted on the [OIII] image.  The same slit position was used to
acquire the H$\alpha$+[NII]658.4 spectrum. The slit width of 0\farcs8
is plotted to scale.

\begin{figure*}
\epsscale{2.0}
\plotone{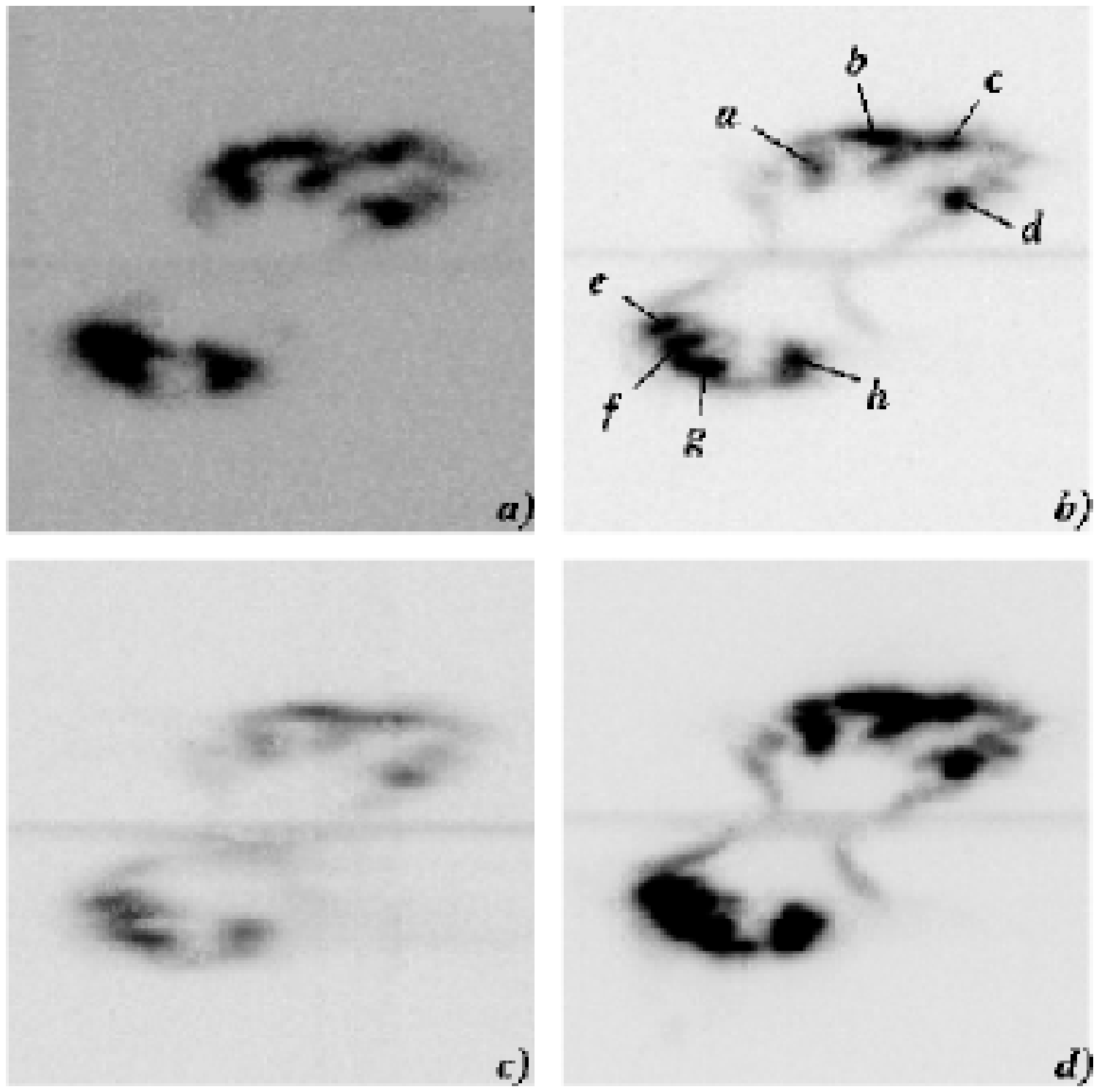}

\caption{The 2--D spectra of Sa2--237. The spatial direction is vertical, the 
spectra run horizontally, and each sub-frame is 0.83nm by
58\arcsec. a) is the [NII]654.8nm line, b) is H$\alpha$, c)
[OIII]500.7nm, and d) [NII]658.3nm.  The letters a--h in the H$\alpha$
spectrum indicate the various parts of the spectrum where we have
extracted velocities by gaussian fits. These features are identifiable
in all four spectral lines. The extracted velocities are listed in
Table~1.\label{fig2}}
\end{figure*}

Additional data for this paper has been extracted from the literature.  
This includes the following fluxes and magnitudes for Sa2--237.

IRAS: 12$\mu$ = 1.17 25$\mu$ = 6.01 60$\mu$ = 16.56 100$\mu$ =
8.29 (Jy); Persi et al. (1987): J=11.85 H=11.05 K=10.72 L=8.7; 2MASS:
J=13.06 H=12.57 K=11.83; Tylenda et al. (1991): B=16.08 V=15.50
\& H$\beta$=2.5$\cdot$10$^{-17}$W$\cdot$m$^{-2}$;
Condon \& Kaplan (1998): 1.4GHz (21cm)=5.8$\pm0.5$ mJy.

The Persi et al. and IRAS results are nearly co-eval (1984-1985) while
the 2MASS magnitudes are from 1998; note the differences of up to 1.5
mag between the NIR measurements over the time interval of 14 yrs,
indicating variability of the source.

The reported systemic radial velocity of Sa2--237 from Beaulieu et al. (1999)
is -81km$\cdot$s$^{-1}$.

\section{Analysis}

Radial velocities were extracted from the more intense regions
of the spectra labeled from a to h in Figure~2.  2--D gaussian
fits were used to determine the velocities listed in Table~1.

\begin{table}[h]
\begin{center}
\caption{Measured line velocities in km$\cdot$s$^{-1}$ with respect to 
the rest $\lambda$ (nm). \label{tbl-1}}
\begin{tabular}{ccccc}

Line & [OIII] & [NII] & H$\alpha$ & [NII]  \\
\tableline
Part &500.685& 654.806 & 656.280 & 658.339 \\
\tableline
a &  54 &  44 &  43 &  43 \\
b &  14 &   0 &   6 &   0 \\
c & -60 & -54 & -48 & -55 \\
d & -60 & -58 & -57 & -59 \\
e & 145 & 153 & 149 & 152 \\
f & 134 & 143 & 142 & 143 \\
g & 129 & 124 & 127 & 126 \\
h &  56 &  58 &  56 &  57 \\
\tableline
2V$_{max}$ & 205 & 211 & 206 & 211 \\

\end{tabular}
\end{center}
\end{table}

2V$_{max}$ is the projected expansion velocity between the fitted
regions with the largest velocity difference.  We are aware that there
is very faint emission at higher velocities but the S\/N ratio is too
low to use these data for velocity calculations. Also, the smearing
out due to seeing dictates the use of the fitted peaks for velocity
determinations. This V$_{max}$ forms a lower limit to the true space
expansion velocity since the projection has not been taken into
account yet.

Note the difference in 2V$_{max}$ between the [NII] lines
(211km$\cdot$s$^{-1}$) and the [OIII] (205km$\cdot$s$^{-1}$) \&
H$\alpha$ (206km$\cdot$s$^{-1}$) lines.  An explanation lies in the
behavior of the material and the excitation lines.  [NII] and other
low excitation lines (e.g. [OI], [SII]) form in the outer regions of
these nebulae, while [OIII] is typically found closer to the central
object.  The velocity of the material tends to be directly
proportional to the distance from the central object, hence the higher
velocities in the [NII] line as shown in Sa2--237. As far as we are
aware, this is the first time that this effect has been observed in a
bipolar nebula; it is common in elliptical PNe.

A small intensity asymmetry between the blue and red shifted loops
in the spectra indicates the likely possibility of a small
misalignment of the slit.  Such an offset would show up
as an intensity asymmetry due to the point symmetric nature of the
nebula.  An estimated offset of 0\farcs5 from the central object is
sufficient to have caused this effect.  Our conclusions, however, are
not affected by this small offset.

We have derived the systemic velocity as the centre of the point
symmetry of the eight shaped velocity--space plot. This follows
directly from our modeling and is an accurate estimate of the system
radial velocity. By applying the appropriate correction we find the
heliocentric radial velocity of Sa2--237 to be -30km$\cdot$s$^{-1}$.
The LSR velocity for Sa2--237 is -16km$\cdot$s$^{-1}$. Note that this
is significantly different from the velocity reported by Beaulieu et
al. 1998 \& Durant et al 1998). For objects at l=11\degr~in the
Galactic plane we expect a small positive velocity of about
V$_{LSR}$=+10km$\cdot$s$^{-1}$ at 2kpc. Our
V$_{LSR}$=-16km$\cdot$s$^{-1}$, possibly indicates somewhat peculiar
kinematics for Sa2--237 but we still place the object in the disk,
albeit a bit far from the plane for a bipolar (type I) PN at
255pc. With the Beaulieu velocity it would have been impossible to
kinematically place Sa2--237 in the disk.  

The fluxes and magnitudes listed in Section~1 are used to create the
energy distribution of Sa2--237 in Figure~3. The NIR values plotted
are those from Persi et al. 1987, since they are nearly co--eval with
the IRAS data.

\begin{figure}[h]
\epsscale{0.8}
\plotone{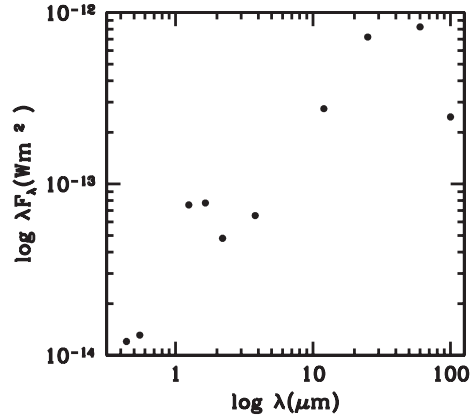}
\caption{The energy distribution of Sa2--237. \label{fig3}}
\end{figure}

There is clearly a major contribution from the IRAS fluxes compared to
the visible and NIR.  This is addressed further below.  Integrating
the {\it fluxes} over wavelength, and correcting by a factor of 1.5
(Meyers et al. 1987) for the unobserved parts of the spectrum, yields
L=77d$^2$L$_{\sun}$ for the luminosity of Sa2--237, where d is the
distance to Sa2--237 in kiloparsecs.

\section{Model and discussion}

Using a simple model for the shape and kinematics of a generic bipolar
nebula, in the manner of Solf \& Ulrich (1985), we fit the morphology
and spectrum of Sa2--237 simultaneously. The model has the following
properties: the velocity of expansion is proportional to the distance
from the center of the material, there is symmetry about the
equatorial plane, and no time evolution. The nebula therefore expands
in a self--similar way. The parameters are: the polar expansion
velocity, V$_p$, the ratio of polar to equatorial velocity, R$_v$, the
inclination to the line of sight of the polar axis, i, and the
age$\cdot$distance$^{-1}$, P$_{ad}$. The expansion velocity as a
parametrized function of latitude in the nebula is given by:\\

V($\phi$) = V$_e$+(V$_p$-V$_e$)sin$^{\gamma}$($\vert\phi\vert$)\\

where $\phi$ is the latitude, V$_e$ \& V$_p$ are the equatorial and
polar expansion velocities respectively, and $\gamma$ is the shape
factor. For more details see the paper by Solf \& Ulrich (1985) whose
parametrization we have followed here.

By fitting the shape of the image and the spectrum simultaneously, we
derive the following parameters for Sa2--237: The inclination, i, to
the line of sight is 70\degr and the maximum expansion velocity is
260km$\cdot$s$^{-1}$.

\begin{figure*}
\epsscale{2.0}
\plotone{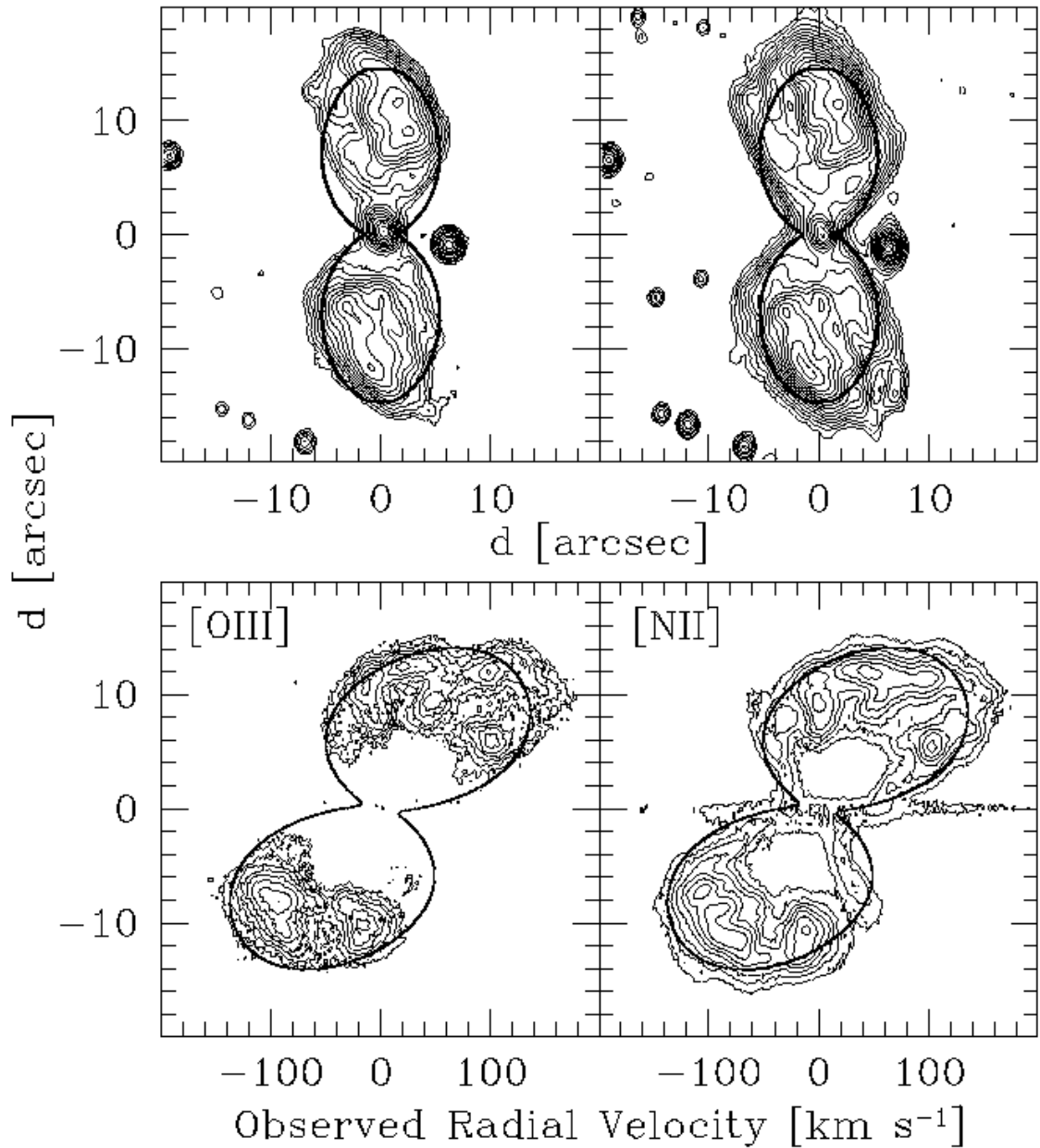}
\caption{Model fits to the images and spectra of Sa2--237. Top: left the 
[OIII] image, right the H$\alpha$ image with model nebula over plotted; 
bottom: left the [OIII] spectrum, right the H$\alpha$ spectrum with model 
spectra over plotted.\label{fig4}}
\end{figure*}

We estimate the distance to Sa2--237 from the observational and
literature data.  Comparision with M2--9, a well studied PN (although
some authors consider it to be a symbiotic nova; Sa2--237 may well be
such a nove as well) which has a similar morphology as Sa2--237 and a
well-determined distance of 640pc (Schwarz et al. 1997), and with
Mira, gives the following results:\\

Scaling the M2--9 distance of to that of Sa2--237 by the object's
1.4~GHz radio fluxes gives 1.66kpc, by luminosity gives 2.68kpc and by
size gives 2.11kpc. A similar comparison with the IRAS flux of o~Cet,
which has an accurate Hipparcos distance of 168pc, and assuming that
the cool component in Sa2--237 is also a Mira, made plausible by the
$\geq$1.5mag variability of Sa2--237, gives 1.96kpc. The average
distance to Sa2--237 based on these four estimates, is 2.1kpc, with a
standard deviation of 0.37kpc, which can be taken as the formal error
on this value. {\it We therefore adopt a distance of 2.1kpc$\pm$0.37kpc to Sa2--237.}

The distance of 2.1kpc is used to compute the following parameters for
Sa2--237. The size of the object is 0.37pc, it's luminosity is
340L$_{\sun}$, and the age of the nebula, assuming constant expansion
velocity, is 624yrs.  Using the model-derived inclination of i=70$^o$,
the space velocity of expansion of Sa2--237 is 308km$\cdot$s$^{-1}$,
using the maximum measured expansion velocity 2V$^{max}$. Note that
this tends to slightly overestimate the space expansion velocity since
the maximum projected velocity, for objects near the plane of the sky,
is not the polar one, but more due to material at an inclination angle
marginally smaller than the 70\degr derived from the model, and
located at a latitude slightly below the poles of the nebula. This
would bring the observed space velocity nearer that derived from the
model (see also below).

Comparing the results derived from the observational data with the
model parameters, we get that the age\/distance parameter,
P$_{ad}$=0.30, in good agreement with P$_{ad}$=0.28 from the model,
given the uncertainties in all parameters. The computed expansion
polar velocity at i=70$^o$ is 308km$\cdot$s$^{-1}$, cf.
260km$\cdot$s$^{-1}$ from the model fitting. Note that only 4$^o$
difference in i, that is i=66$^o$, gives 260km$\cdot$s$^{-1}$ instead
of 308km$\cdot$s$^{-1}$, making also this a reasonable
agreement. {\it We therefore adopt a space expansion velocity of $\leq$308km$\cdot$s$^{-1}$ 
for Sa2--37.}

All these parameters are typical of bipolar nebulae, and this
strengthens our confidence that Sa2--237 is indeed a bipolar PN or
symbiotic nova such as M2--9, He2--104, or BI~Cru (Corradi \& Schwarz
(1993) and various others.

The orientation near the plane of the sky of Sa2--237 implies that the
central object is likely obscured by equatorial material, which would
cause a large extinction. There must be a pole to equator density
gradient to produce the bipolar shape, with the highest density at the
equator. This is the reason why the IRAS contribution is much larger
than the optical+NIR flux, as the FIR radiation comes from the
equatorial disk as reprocessed optical and NIR flux absorbed in the
torus. The energy distribution of M2--9 is similar and this object is
also near the plane of the sky. Note that radio fluxes are not (or
much less) affected by this and the distance ratios computed from the
radio and IRAS fluxes confirm this: the radio distance is smaller
(1.66kpc) than the IRAS ($\approx$ luminosity) distance (2.68kpc),
indicating some extinction in the IRAS flux but none in the radio.

One can, over the next years, attempt to measure the expansion parallax for
Sa2--237, as we have done for M2--9, as it should be 0\farcs053
yr$^{-1}$, at our distance of 2.1kpc. Taking one or two images per year would
give a result within a few years, and would allow a direct check on
our derived distance. We have started such a program of observations.

In the images, the nucleus of the nebula is relatively bright, but in
the spectra no strong H$\alpha$ emission appears. It is likely that it
is a stellar spectrum, not an emission line core, as in symbiotic
stars. It would be interesting to take a spectrum of this star and
attempt to classify it.

\section{Summary \& conclusion}

We summarize the properties of Sa2--237. 

It is a bipolar planetary nebula, similar to M2--9, M1--16 and others,
projecting 34\arcsec onto the sky, at a distance of 2.1kpc, with a
size of 0.37pc, L=340L$_{\sun}$, and expanding at
2V$_{exp}$=616km$\cdot$s$^{-1}$. It's point symmetric morphology hints at
the presence of precession and thus the central object may well be a
binary. The object's inclination is 70\degr to the line of sight.

Based on the symmetry in the observed expansion velocities, we adopt a
heliocentric radial velocity for Sa2--237 of -30km$\cdot$s$^{-1}$, and
place it in the Galactic disk at z=255pc.

We suggest that the expansion parallax can be measured on a relatively
short time scale as it is $\approx$0\farcs05 p.a. This would confirm
or adjust the distance to Sa2--237.

Spectra of the central object may tell us what properties it has, as
it seems to be stellar in nature and not an emission line core, as in
many other such objects.

\section{Acknowledgments}

The CTIO Research Experiences for Undergraduates Program is funded by the 
National Science Foundation.

\end{document}